\documentclass[twocolumn]{aastex6}

\def\gtsima{$\; \buildrel > \over \sim \;$}
\def\ltsima{$\; \buildrel < \over \sim \;$}
\def\gsim{\lower.5ex\hbox{\gtsima}}
\def\lsim{\lower.5ex\hbox{\ltsima}}

\shorttitle{Reverberation in H1743-322}
\shortauthors{De Marco and Ponti}

\begin{document}


\title{The reverberation lag in the low mass X-ray binary H1743-322}


\author{Barbara De Marco and Gabriele Ponti}
\affil{Max-Planck-Institut f\"{u}r Extraterrestrische Physik \\
Giessenbachstrasse 1 \\
D-85748, Garching, Germany}








\begin{abstract}

The evolution of the inner accretion flow of a black hole X-ray binary (BHXRB) during the outburst is still a matter of active research. X-ray reverberation lags are powerful tools to constrain the disk-corona geometry. We present a study of X-ray lags in the black hole transient H1743-322.
We compared results obtained from the analysis of all the publicly available XMM-Newton observations. These observations were carried out during two different outbursts, happened in 2008 and 2014. 
During all the observations the source is caught in the hard state and at similar luminosities ($\mathrm{L_{3-10\ keV}/L_{Edd}\sim 0.004}$). We detected a soft X-ray lag of $\sim$60 ms, most likely due to thermal reverberation.
We did not detect any significant change of the lag amplitude among the different observations, indicating a similar disk-corona geometry at the same luminosity in the hard state. 
On the other hand, we observe significant differences between the reverberation lag detected in H1743-322 and in GX 339-4 (at similar luminosities in the hard state), which might indicate variations of the geometry from source to source.

\end{abstract}

\keywords{X-rays: binaries -- X-rays: individual (H1743-322) -- accretion, accretion disks}



\section{Introduction} 
\label{intro}

Accreting black hole (BH) systems emit variable X-ray radiation from the inner comptonizing region (corona), that illuminates the surrounding matter (e.g. Shapiro, Lightman, \& Eardly 1976; Haardt \& Maraschi 1991, 1993). 
The dense material of the optically thick disk is expected to respond quasi-instantaneously to X-ray irradiation (Guilbert \& Rees 1988). Therefore, the relevant 
time delay between the primary and the reprocessed emission is the light crossing time between the two emitting regions. 
This delay, so-called `X-ray reverberation lag', is a powerful tracer 
of the geometry of the innermost accretion flow (e.g. Uttley et al. 2014 and references therein).

X-ray reverberation is now commonly observed in radio quiet active galactic nuclei (AGN; Fabian et al. 2009; De Marco et al. 2011; Zoghbi et al. 2012; De Marco et al. 2013a; Kara et al. 2013a). 
In these sources the reverberation lag scales with the black hole mass, $\rm{M_{BH}}$,  
mapping distances of only a few gravitational radii, $\rm{r_{g}}$ (De Marco et al. 2013a; Kara et al, 2013a; Uttley et al. 2014).
This is expected if the corona is compact and the disk extends down to very short distances from the supermassive black hole (SMBH; e.g. Wilkins \& Fabian 2013).

Black hole X-ray binaries (BHXRBs) are thought to be the stellar-mass counterpart of SMBHs (e.g. Czerny et al. 2001; K\"{o}rding et al. 2007). 
Indeed, many of their observational properties resemble those seen in AGN, once the difference in $\rm{M_{BH}}$ is accounted for (e.g. Fender et al. 2006). 
However, BHXRBs show a complex evolutionary pattern throughout the outburst (e.g. Fender et al. 2004; Belloni et al. 2005; Dunn et al. 2010; Mu\~noz-Darias et al. 2011). 
In particular, these sources undergo substantial variations of their spectral and timing properties. These are usually explained in terms of variations of the truncation radius of the optically thick and geometrically thin disk (Esin et al. 1997; Done, Gierli\'nski \& Kubota 2007). Currently, observational evidences exist both in favour and against this scenario (e.g. Miller et al. 2006; Kolehmainen et al. 2014; Petrucci et al. 2014; Garc\'ia et al. 2015; Plant et al. 2015; Basak \& Zdziarski 2016), so that the way the standard thin disk evolves during the outburst is still debated.
Reverberation lags add a further dimension to studies of the inner structure of the flow, and can be used to put firm constraints on its geometry (De Marco et al. 2013a; Uttley et al. 2014; De Marco et al. 2015a, hereafter DM15).\\
When compared to those seen in AGN, reverberation lags detected in the hard state of BHXRBs are about one order of magnitude larger than expected from a scale-invariant
disk-corona geometry (Uttley et al. 2011; De Marco et al. 2013b; DM15). These values are indicative of a different geometry characterizing the hard state of 
BHXRBs, and are in agreement with the predictions of truncated disk models (e.g. Esin et al. 1997). In addition, a decrease of the reverberation lag amplitude as a function of luminosity throughout the hard state is observed in GX 339-4 (DM15). This is expected if the disk truncation radius gradually decreases as the source approaches the transition to disk-dominated soft states (Done, Gierli\'nski \& Kubota 2007).\\
Because of the lack of suitable XMM-Newton observations, the evolution of the reverberation lag during the outburst could be studied only in GX 339-4 (DM15). 
However, signatures of reverberation have been observed also in one observation of H1743-322, carried out during the 2008 outburst.\\
H1743-322 is a transient system, discovered by the \emph{Ariel V} satellite (Kaluzienski \& Holt 1977), with X-ray spectral and timing properties typical of a BHXRB (e.g. McClintock et al. 2009; Motta et al. 2010). A large-scale radio and X-ray jet associated with this source was observed during the 2003 outburst. By modeling its trajectory, Steiner et al. (2012) estimated a distance of $8.5\pm0.8$ kpc. Moreover, the inclination angle of the jet ($i=75^{\circ}\pm3^{\circ}$; Steiner et al. 2012) and the presence of X-ray dipping episodes (Homan et al. 2005; Shidatsu et al. 2014) suggest H1743-322 to be a high inclination system. This is also confirmed by the detection of equatorial disk winds during the soft state of the source (Ponti et al. 2012; 2015) as well as the shape of its hardness-intensity diagram (HID; Mu\~noz-Darias et al. 2013). 
Currently, there are no dynamical constraints on the BH mass. A BH mass of $13.3\pm3.2\rm{M_\odot}$ has been inferred by Shaposhnikov \& Titarchuk (2009) from the scaling of the correlation patterns of the source's spectral and timing characteristics.\\
H1743-322 displays frequent outbursts. XMM-Newton has recently re-observed it three times, during the 2014 outburst. 
Both the 2008 and 2014 are ``failed outbursts'' (Capitanio et al. 2009; Stiele \& Yu 2016), meaning that the source did not reach the transition to the soft state and remained in the hard state for the entire outburst.\\
In this paper we report on the study of all the available XMM-Newton observations of H1743-322 during outburst.
We aim to study the evolution of the X-ray reverberation lag in this source, and, by comparison with the evolution observed in GX 339-4, to test whether, in a given accretion state, the disk-corona geometry is the same for different sources.

\begin{figure*}
\centering
\vspace{2.5cm}
\begin{tabular}{p{10cm}}
\includegraphics[height=0.68\textwidth,width=0.55\textwidth,angle=0]{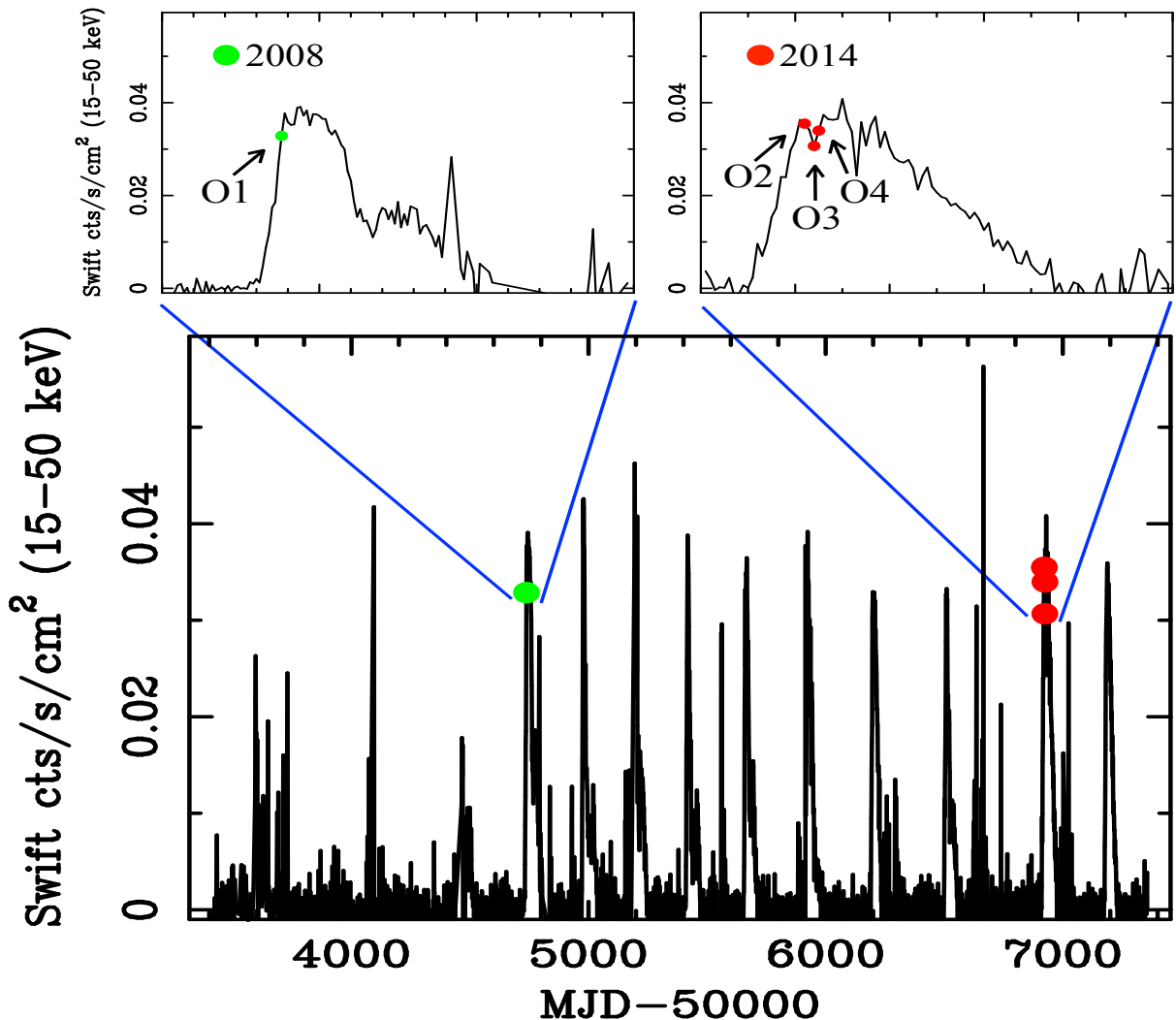}\\ 
\end{tabular}
\caption{Swift/BAT long term light curve of H1743-322. The green and red dots mark the positions of the XMM-Newton observations analysed in this paper. The two small panels show a zoom of the 2008 and 2014 outburst lightcurves.}
\label{fig:lc}
\end{figure*}

\section{XMM-Newton observations and data reduction}
We analysed the three XMM-Newton observations of H1743-322 carried out during the 2014 outburst. The available data set spans a time 
interval of about four days. 
In addition we re-analysed the 2008 September 29 observation (ObsID 0554110201), already presented in DM15. The log of the observations is reported in Table \ref{tab:ObsLog}. Hereafter we will refer to each observation using the nomenclature reported in column (3) of Table \ref{tab:ObsLog}.\\
We primarily used EPIC-pn (Str\"uder et al. 2001) data in Timing mode. However, for the spectral analysis presented in Sect. \ref{sec:analysis} we also included RGS (den Herder et al. 2001) data, in order to better constrain the soft X-ray band ($\rm{E}\leq 2$ keV) continuum.  
Data reduction was carried out following standard procedures (e.g. DM15). We used the XMM Science Analysis System (SAS v14) 
and calibration files as of 2015 December.\\
The effective exposures (after removal of background flares) of each observation are reported in Table \ref{tab:ObsLog}. EPIC-pn source and background counts were extracted in the interval RAWX: 29-47 and RAWX: 58-64, respectively. We checked for the presence of pile-up using the SAS task \emph{epatplot}. Beside distortions at $\rm{E}\lsim 0.5$ keV due to low-energy electronic noise (Guainazzi et al. 2010), we did not detect any significant deviation of the pattern-fraction distribution from the theoretical curves, which would indicate the presence of pile-up. 
We verified that the background does not have significant effects on the lag-energy spectra presented in Sect. \ref{sec:lags}. Thus, in order to retain a higher signal-to-noise ratio, we decided to not perform any background subtraction for the timing analysis.\\

\section{Analysis}
\label{sec:analysis}
Fig. \ref{fig:lc} shows the long-term 15-50 keV light curve of H1743-322 as measured by the Swift/BAT Hard X-ray transients monitor (Krimm et al. 2013). The source is characterized by an intense activity, displaying about one outburst per year.
The position of the XMM-Newton observations analysed in this paper are marked in this plot. 
All the analysed XMM-Newton observations caught the source during the increasing phase of the outburst, at similar hard X-ray fluxes.

We simultaneously fit the EPIC-pn and the combined RGS1 and RGS2 spectra of each observation (the spectra of O2 are shown in Fig. \ref{fig:specs}). We used a model comprising a multi temperature disk black body, a Comptonization component, and a Gaussian line to account for Fe K emission (\emph{tbabs*[diskbb+nthcomp+gau]} in Xspec; Wilms et al. 2000; Zdziarski et al. 1996; Zycki et al. 1999). The high-energy cutoff of the Comptonization component was fixed at 100 keV, while the seed photon temperature was tied to the inner disk temperature. The galactic column density $\rm{N_H}$ was left free to vary.\\
As more extensively discussed in Sect. \ref{sec:discussion}, the EPIC-pn low-energy effective area might be affected by uncertainties in the model used to describe the distortions of the spectrum due to incomplete charge collection (Popp et al. 1999; 2000). For this reason, in the fits we ignored energies below 1.5 keV, and used RGS data in the range 0.8-2 keV to constrain the soft X-ray band continuum. The parameters of the best-fit model are listed in Table \ref{tab:fits}. The best-fit galactic column density is $\rm{N_H}\sim 2\times10^{22}\rm{cm}^{-2}$, consistent with values reported in the literature (e.g. Stiele \& Yu 2016; Shidatsu et al. 2014), though slightly higher than derived from galactic $\rm{HI}$ surveys ($\rm{N_H}\sim 7\times10^{21}\rm{cm}^{-2}$, e.g. Dickey \& Lockman 1990), thus suggesting local absorption. The fits yielded best-fit values of the disk inner temperature $\rm{T_{in}}\sim 0.3-0.4$ keV. The disk-to-power law flux at $\sim$ 1 keV is consistent with being of the order of 30-40 percent during all the observations.

Overall we obtain satisfactory fits (see Table \ref{tab:fits}), but the extrapolation of the best-fit model to energies $\lsim 1$ keV leaves significant residuals in the EPIC-pn spectra, which are not present in the RGS (Fig. \ref{fig:specs}). Indeed, the RGS flux is observed to drop by a factor $\sim$ 100 between 1.5 keV and 0.8 keV, while the drop seen in the EPIC-pn data is only a factor $\sim$ 20. Similar soft excess residuals have been observed in other sources (e.g. Boirin et al. 2005; Martocchia et al. 2006; Sala et al. 2008; Hiemstra et al. 2011). We discuss the origin of these residuals in Sect. \ref{sec:discussion}.

\subsection{Accretion state}
\label{sec:acc_state}
We used both spectral and timing indicators to determine the accretion state of H1743-322 during all the analysed observations. 
We measured the spectral hardness from our best-fit model (Sect. \ref{sec:analysis}) as the ratio between fluxes in the 6-10 keV and the 3-6 keV energy bands (see Table \ref{tab:fits}). We found the hardness ratios to be in the range $\sim 1.15-1.18$, not showing significant spectral variability among the different observations. 
In the hardness-intensity diagram of the source (e.g. Dunn et al. 2010) these values are typical of the hard state, as also shown by Stiele \& Yu (2016) for O4.\\
In Table \ref{tab:fits} we also report the total 3-10 keV flux ($\rm{F}_{3-10\ keV}$) of the source. The X-ray flux during the 2014 observations is only slightly higher (by a factor $\sim 1.2$) than during O1, while it is characterized by similar values among the 2014 data sets.

As a model independent method for determining the accretion state of the source, we estimated its power spectral density function (PSD). The PSD shows band limited broad band noise and type C quasi periodic oscillations (QPO, the main harmonic being detected at $\nu\sim 0.2$ Hz) during all the observations. These are all features typical of the hard state and were reported also by Stiele \& Yu (2016) for O4. The characteristic frequencies appear constant among all the observations (including O1), besides a slight increase (a factor $\sim$1.15) of the frequencies of the QPO and of the high frequency break during O3 and O4.\\
Finally, we computed the fractional root-mean-square variability amplitude ($\rm{F_{var}}$). Following Mu\~noz-Darias et al. (2011), we measured $\rm{F_{var}}$ in the 2-10 keV energy band and over the 0.1-50 Hz frequency range. The values we obtained (see Table \ref{tab:fits}) indicate high variability and again point to the source being in the hard state (Mu\~noz-Darias et al. 2011; Heil et al. 2012; De Marco et al. 2015b) during all the analysed observations.

\subsection{Lag-energy and covariance spectra}
\label{sec:lags}
We extracted EPIC-pn light curves with a time resolution of 0.02 s in adjacent energy bins and a broad reference band. As the reference band we used the 0.5-10 keV energy band\footnote{Since the lag-energy spectrum may depend on the choice of the reference band (Uttley et al. 2014) we cross-checked our results using the power law-dominated, narrower band 1.5-10 keV as the reference.}.
The light curves were divided into segments of $\sim$ 41 s length, and the cross-spectrum computed from each segment. We then averaged over the different cross-spectra and derived an estimate of the time lag between each bin and the reference band, integrating over given frequency intervals. To remove the contribution of cross correlated Poisson noise, for each energy bin, we discarded counts from that bin when computing the light curve of the reference band. 
For more details about the procedure we refer to Uttley et al. (2014).

\begin{figure}
\centering
\vspace{2.5cm}
\begin{tabular}{p{10cm}}
\includegraphics[width=0.45\textwidth,angle=0]{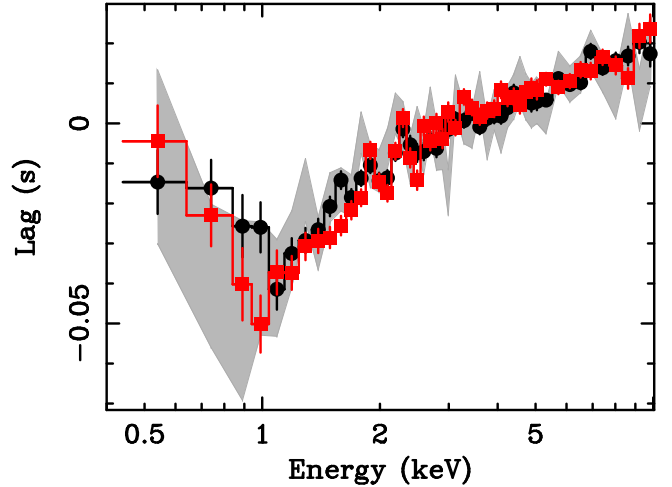}\\ 
\end{tabular}
\caption{Lag-energy spectra of H1743-322 in the frequency interval 0.1-1 Hz. The black dots and red squares refer, respectively, to O2 and O3$+$O4, and correspond to the observations taken during the 2014 outburst. The gray contours refer to O1, corresponding to the observation taken during the 2008 outburst.}
\label{fig:lagE}
\end{figure}

In Fig. \ref{fig:lagE} we report the lag-energy spectra in the frequency range 0.1-1 Hz. It is worth noting that at frequencies $\lsim$ 0.1 Hz the lag-energy spectra in the soft band are much noisier, as a consequence of a drop of coherence (from $\sim$1 to $\sim$0.5) below $\sim 2$ keV. On the other hand, at frequencies $\gsim$ 1 Hz, the lag-energy spectra are Poisson noise-dominated, particularly in the soft band, where the count rate is lower due to galactic absorption. 
Overplotted in Fig. \ref{fig:lagE} are the $1\sigma$ contours of the lag-energy spectrum as measured from O1 (see also DM15) within the same frequency range. All the spectra show the presence of soft band delays (at $\rm{E}\lsim 1$ keV) superimposed upon the hard lags observed at higher energies (associated with the power law component). 
The new XMM-Newton observations, being longer than O1, allowed us to significantly reduce (by a factor $\sim$ 3) the errors on the lag measurements, thus increasing the significance of the soft lag detection and characterization.

\begin{figure}
\centering
\vspace{2.5cm}
\begin{tabular}{p{10cm}}
\includegraphics[height=0.27\textwidth,angle=0]{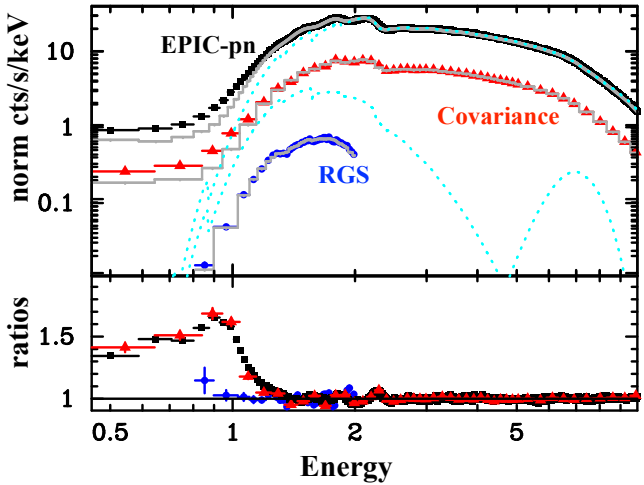}\\ 
\end{tabular}
\caption{The plot shows the EPIC-pn and RGS flux-energy spectra, and the covariance spectrum of O2 (black squares, blue dots, and red triangles respectively). The dotted (light blue) curves are the single components of the best-fit model (Sect. \ref{sec:analysis} and Table \ref{tab:fits}) rescaled for the EPIC-pn effective area. The EPIC-pn flux-energy and covariance spectra are fitted in the energy range 1.5-10 keV. The extrapolation of the best-fit model (convolved for the response matrix, gray curves) down to 0.5 keV is shown to highlight the soft X-ray band residuals. The ratios to the best-fit model are shown in the lower panel.}
\label{fig:specs}
\end{figure}

To study the origin of the soft lags detected in H1743-322 we measured the 0.1-1 Hz covariance spectra, and fit those at energies $\rm{E}\geq 1.5$ keV with the same model used for the flux-energy spectra (Sect. \ref{sec:analysis}). All the parameters of the model were tied to those of the flux-energy spectrum of the corresponding observation, with the exception of the normalization of each component and the spectral index of the power law. In Fig. \ref{fig:specs} we show the covariance spectrum and the ratio to the best-fit model for observation O2 (red triangles). The model describes the data well in all the observations. However, the extrapolation of the best-fit model to low energies ($\rm{E}\leq1.5$ keV) leaves excess residuals similar to those observed in the EPIC-pn flux-energy spectra. We discuss this result in Sect. \ref{sec:discussion}.

\begin{figure}
\centering
\vspace{2.5cm}
\begin{tabular}{p{10cm}}
\includegraphics[width=0.45\textwidth,angle=0]{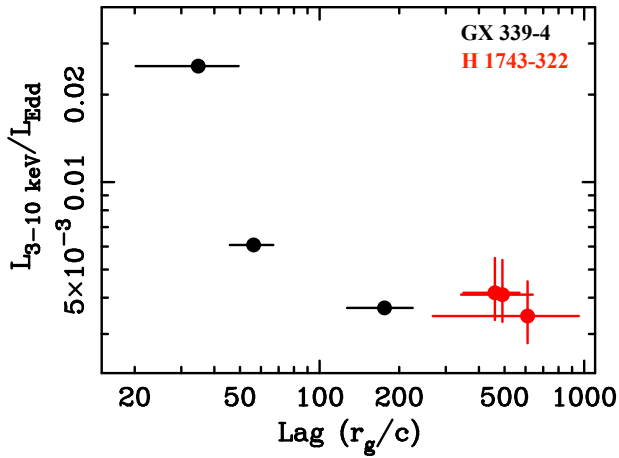}\\ 
\end{tabular}
\caption{The reverberation lag amplitude observed in the hard state of H1743-322 (this paper) and GX 339-4 (DM15) as a function of Eddington-scaled 3-10 keV X-ray luminosity (respectively red and black dots). The errors on the luminosity of H1743-322 account for the uncertainty on the distance and the BH mass of the source. For GX 339-4 a fixed value of $\rm{M}=8 \rm{M_{\odot}}$ and $\rm{d}=8$ kpc is assumed.}
\label{fig:L_vs_lag}
\end{figure}

\begin{table}
\caption{The table reports: (1) observation ID of the analysed XMM-Newton observations; (2) the date of the observation; (3) nomenclature used throughout the paper; (4) the net exposure after removal of proton flares.}
\label{tab:ObsLog}
\centering
\vspace{0.2cm}

\begin{tabular}{c c c c}
\hline
	(1)                & (2)                     & (3)        & (4)    \\
      Obs ID          & Date                  & Obs\#  & Net Exp    \\
                           & [yyyy-mm-dd]    &             & [$\rm{ks}$] \\
\hline			                     
      0554110201  & 2008-09-29       &  O1      & 21          \\
      0724400501  & 2014-09-21       &  O2      & 137        \\
      0724401901  & 2014-09-23       &  O3      & 79          \\
      0740980201  & 2014-09-24       &  O4      & 49         \\

\hline

\end{tabular}
\end{table}

\begin{table*}
\caption{Best-fit model obtained by simultaneous fit of the EPIC-pn spectra in the range $\rm{E}=1.5-10$ keV and combined RGS spectra in the range $\rm{E}=0.8-2$ keV. The table reports: (1)-(5) the best-fit values of the relevant parameters; (6) the 3-10 keV flux; (7) the disk-to-power law flux at 1 keV; (8) the hardness ratio; (9) the obtained $\chi^2$ statistics; (10) the 2-10 keV fractional root-mean-square variability amplitude in the 0.1-50 Hz frequency range. Errors are quoted at 90 percent confidence for 1 parameter of interest ($\Delta\chi^2=2.7$).}
\label{tab:fits}
\centering
\vspace{0.2cm}
\begin{tiny}
\begin{tabular}{c c c c c c c c c c c}
\hline       
   & 	(1)         &  (2)  & (3) & (4) & (5) & (6) & (7) & (8) & (9) & (10) \\
     &  $\rm{N_{H}}$             & $\Gamma$   & $\rm{T_{in}}$ & $\rm{E_{gau}}$ & $\sigma_{gau}$ & $\rm{F_{3-10\ keV}}$ & $\rm{F_{disk}/F_{pow}}$ & $\rm{F_{6-10\ keV}}/\rm{F_{3-6\ keV}}$ &  $\chi^2/\rm{dof}$ & $\rm{F_{var\ 2-10 keV}}$ \\
	    &  $10^{22} \rm{cm^{-2}}$ &                     &   keV              &        keV            & keV                   &  $10^{-10} \rm{erg/s/cm^2}$ &              &            &         &    \\

\hline			                     

O1      &   $2.17\pm0.06$ &  $1.47\pm0.01$ &  $0.29\pm0.01$ & $6.83\pm0.20$  & $0.6^{fixed}$               &   $6.93\pm0.08$   &  $0.62\pm0.27$  & $1.15\pm0.02$ & 2145/2056 & $0.355\pm0.003$\\
O2      &   $2.06\pm0.03$ &  $1.43\pm0.01$ &  $0.37\pm0.01$ & $7.16\pm0.11$ & $0.87^{+0.13}_{-0.10}$ & $8.21\pm0.09$ & $0.35\pm0.08$ & $1.18\pm0.02$ &  4576/2333 & $0.339\pm0.002$ \\
O3      &   $2.03\pm0.03$ &  $1.43\pm0.01$ &  $0.41\pm0.02$ & $6.92\pm0.09$ & $0.63\pm0.08$  &   $8.30\pm0.09$  & $0.33\pm0.11$  & $1.17\pm0.02$ &    3511/2253 & $0.320\pm0.002$\\
O4      &   $1.99\pm0.03$ &  $1.41\pm0.01$ &  $0.47\pm0.03$ & $6.73\pm0.08$ & $0.47\pm0.11$  &   $8.37\pm0.09$ & $0.26\pm0.10$   & $1.18\pm0.02$ & 3319/2186 &$0.317\pm0.002$ \\
\hline

\end{tabular}
\end{tiny}
\end{table*}

\section{Discussion}
\label{sec:discussion}
The detection of a soft band delay in H1743-322 was first reported in DM15 in a single XMM-Newton observation of the 2008 outburst (O1), during which the source was in the hard state. 
During the 2014 outburst the source has been caught by XMM-Newton again in the hard state, and at similar luminosity as during O1 (Sect. \ref{sec:acc_state}). 
Here we report the detection of a soft band ($\rm{E}\lsim 1$ keV) time lag also in the latest observations (Sect. \ref{sec:lags} and Fig. \ref{fig:lagE}).\\
Given the relatively high absorption column density ($\rm{N_H\sim 2\times10^{22}\ cm^{-2}}$) the soft X-ray band below $\sim$0.8 keV should be totally absorbed. However, at these energies both the EPIC-pn flux-energy spectra and the covariance spectra of all the observations show excess residuals, which are not observed in the RGS data (Sects. \ref{sec:analysis} and \ref{sec:lags}, and Fig. \ref{fig:specs}). We discuss here the origin of the excess emission in the soft X-ray band of the EPIC-pn spectrum (and covariance), and of the associated time lag.\\
{\bf \emph{Foreground point sources:}} all the analysed observations were carried out in Timing mode, meaning that all the data along the RAWY axis of one CCD chip are collapsed into a one-dimensional row. It is thus possible that the observed soft X-ray excess is due to the summed contribution of foreground point sources. However, this hypothesis is highly unlikely for at least two reasons. Firstly, the presence of this feature also in the covariance spectra implies that the flux variability of the soft X-ray excess is correlated with the variability of the reference band\footnote{Note that a soft X-ray excess in the covariance spectra is also observed when the power law-dominated 1.5-10 keV band is used as the reference.}. This excludes contribution from uncorrelated nearby sources, and hints to this feature being associated with the source itself. Secondly, we checked for the presence of soft X-ray bright point sources in the field of view of H1743-322 by inspecting the 2010 October 9 XMM-Newton observation (ObsID 0553950201), taken in SmallWindow mode, when the source flux was much lower ($\rm{F_{3-10\ keV}\sim 1.17\times10^{-11}\ erg/s/cm^{2}}$, i.e. a factor $\sim$60-70 fainter than during the observations analysed in this paper). We identified 2 additional point sources $\sim$75 and 134 arcsec away from H1743-322. Assuming their flux is stable, we estimated that their contribution to the 0.5-1 keV total counts of the 2008 and 2014 observations is negligible ($\lsim$1-0.5 percent).\\
{\bf \emph{Dust scattering halo:}} another possibility is that the low-energy excess is due to the contribution of the dust scattering halo (e.g. Xu, McCray, \& Kelley 1986). We note, however, that unless the spatial distribution of the dust is highly inhomogeneous, the scattered spectrum is expected to be absorbed by the same column density as that absorbing the intrinsic spectrum of the source (e.g. Predehl \& Schmitt 1995). Therefore, the dust scattering halo should not give significant contribution in the soft X-ray band below $\sim$0.8 keV. In addition, we notice that in the presence of a variable X-ray source, the scattered signal will arrive with a time delay which depends on the distance of the source and of the dust layer to the observer, as well as on the observed off-axis angle of the scattering site (e.g. Heinz et al. 2015; Vasilopoulos \& Petropoulou 2016). Using eq. 3 of Vasilopoulos \& Petropoulou (2016), and assuming a distance of 8.5 kpc (Sect. \ref{intro}) and $\geq$ 4 kpc respectively for the source and the dust layer, we estimate that only dust located very close to the line of sight (with an off-axis angle of a few fractions of an arcsec) can produce time delays of the order of tens of msec as observed in the analysed data sets. 
The scattered spectrum associated with clouds close to the line of sight to the source (thus able to produce the observed lags) and the intrinsic source spectrum are likely to be absorbed by the same $\rm{N_H\sim 2\times10^{22}\ cm^{-2}}$. This suggests that the dust scattering halo gives negligible contribution to the observed lag.\\
{\bf \emph{Instrumental effects:}} the fact that soft X-ray excess residuals do not seem to be present in the RGS data indicates that instrumental effects might be at play. We find the most plausible explanation to be related to incomplete charge collection for pn-CCDs, which causes the fraction of collected charge to decrease when the photon is absorbed close to the detector surface. This effect is significant for low energy photons ($\rm{E}\lsim 1$ keV), and is responsible for a distortion of the spectrum: in the case of monochromatic light, it leads to the formation of a shoulder and a flat-shelf to the low-energy side of the Gaussian peak (Popp et al. 1999; 2000). In the case of broad band emission, the net effect is that photons of energy $\sim$ 1 keV are transferred to lower energies forming a soft X-ray flat continuum. As a consequence, given that significant emission from the disk component ($\sim$30-40 percent of the power law flux, Table \ref{tab:fits}) is observed at $\sim$ 1 keV, we infer that the soft X-ray excess should include a significant fraction of disk photons.\\
The low-energy response of the detector depends on the calibration of the intensity of the shoulder and the flat-shelf relative to the Gaussian peak. This is currently uncertain, in particular for the Timing mode (F. Haberl, private communication). Thus, we infer that an overestimate of the relative intensity of the spectral features produced by the charge loss process might be the cause of the soft X-ray residuals seen in H 1743-322. Nonetheless this effect does not have any influence on the measured time lag (the lag amplitude is, indeed, a factor $\sim$ 2000 longer than the frame time in Timing mode).\\
{\bf \emph{Thermal reverberation:}} we are left with the hypothesis that the soft lag is intrinsic to the source. Soft lags in the hard state of BHXRBs can be the signature of reverberation of the hard X-ray photons, which irradiate the inner radii of the optically thick disk and are thermalized (Uttley et al. 2011, DM15). According to our best-fit model, the disk contributes about 30-40 percent of the power law flux at $\sim$ 1 keV (Table \ref{tab:fits}; see also Fig. \ref{fig:specs}), where the soft lag starts to be observed. This is in agreement with the values derived for GX 339-4 (DM15). 
Hereafter we consider the physical implications of this interpretation. 

Reverberation lags measure the light crossing time of the distance between the X-ray emitting region (corona) and the optically thick reprocessing region (disk). In the lag-energy spectra the reverberation component must be disentangled from the underlying hard lags, which are ascribed either to propagation of mass accretion rate fluctuations (e.g. Kotov et al. 2001; Arevalo \& Uttley 2006) or to delays related to the Comptonization process (Uttley et al. in preparation), and show a log-linear trend as a function of energy.
In DM15, as a parametrization of the intrinsic reverberation lag amplitude, we used the maximum intensity of the soft band residuals obtained from the extrapolation of the log-linear model fitting the high-energy hard lags. For consistency, we use this same parametrization for H1743-322\footnote{The hard lags of H1743-322 show a double-slope log-linear trend, with a turnover at about 2 keV, similar to that observed in the low frequency ($\lsim$ 0.8 Hz) lag-energy spectra of GX 339-4. To account for this we limit the fit with a log-linear model to the energy range 1-2 keV.}. 
With this method we found the reverberation lag amplitude to be $\tau=64\pm18$ ms and $\tau=60\pm13$ ms for O2 and O3$+$O4, and $\tau=80\pm44$ ms for O1. 
Using the equation of the isodelay surface $\tau=(1+cos\theta)d/c$ (where $d$ is the distance of the reprocessing site from the driving X-ray source and $\theta$ its angle measured from the line of sight to the observer, e.g. Peterson 2001) we inferred an order of magnitude lower limit on the distances involved. We found that the measured lags (considering also their associated errors) give lower limits on the light-crossing distances of $\gsim 250\ \rm{r_{g}}$ and $\gsim 500\ \rm{r_{g}}$ (where we assumed the $\rm{M_{BH}}$ and distance of the system reported in Sect. \ref{intro}), respectively for a source of hard X-rays located above the disc ($\theta=0^{\circ}$) and  for a central source of hard X-rays and a truncated-disk geometry ($\theta=90^{\circ}$). It is important to stress that these are back-of-the-envelope estimates, and that a detailed modelization of both the hard and the reverberation lags in the lag-energy spectra is needed to more precisely constrain these distances (see also discussion below).\\
The lags show little scatter and within the errors they are all consistent with each other. The small observed scatter indicates that the disk-corona geometry is very similar among the different observations. Note that for an optically thick and geometrically thin disk the viscous time scale at radii $\gsim 250\ \rm{r_{g}}$ is $\gsim$6 days (assuming viscosity parameter $\alpha=0.01$). This is longer than the separation among O2, O3, and O4, thus, in a truncated-disk scenario significant variations of the disk inner radius are indeed not expected on the timespan of the latest observations.
Given that the source is always in the hard state (and at similar X-ray luminosities) during all the analysed observations (Sect. \ref{sec:acc_state}), this suggests a similar geometry of the accretion flow associated with a given luminosity in the hard state, even during different outbursts. 
This is in line with spectral studies, that show that the properties of a transient source appear very similar from outburst to outburst.
However, it is worth noting that these observations all correspond to hard states in the initial phase of the outburst. It is yet to be verified whether this is the case also when comparing the same hard state during the rising and the descending phases of the outburst.\\

In Fig. \ref{fig:L_vs_lag} we plot the reverberation lag amplitude (in units of $\rm{r_g}/c$) as a function of 3-10 keV Eddington-scaled luminosity for both H1743-322 and GX 339-4 (as measured in DM15) in the hard state, assuming a source of hard X-ray photons located above the disc and without accounting for the inclination of the system. 
When comparing Eddington-scaled luminosities of different sources, the uncertainty on the distance and the BH mass must be taken into account. 
Here, for H1743-322 we used the distance and the BH mass (and associated errors) reported in Sect. \ref{intro}, while for GX339-4 a distance of 8 kpc and a mass of 8$\rm{M_\odot}$ are assumed. 
From this plot we notice that the reverberation lag measured in H1743-322 is about 2-3 times longer than the lag measured in GX 339-4, at the same luminosity. 
There are many mechanisms, such as lag dilution, inclination, etc., that 
could induce a difference in the measured lag, given the same disk-corona geometry. Alternatively, the offset could be intrinsic. Here below we discuss some of these possible origins.\\
{\bf \emph{Dilution:}} we consider the hypothesis that the disk-corona geometry of H1743-322 and GX 339-4 is the same when the two sources are in the hard state at the same luminosity, but the measured amplitude of the reverberation lag appears different as a consequence of lag dilution effects (e.g. Wilkins \& Fabian 2013; Kara et al. 2013b; Cackett et al. 2014; Uttley et al. 2014). 
Dilution is due to the contribution of the driving continuum in the energy band of the disc component, thus reducing the measured amplitude of the thermal reverberation lag. 
The amount of dilution depends on the disk flux fraction relative to the primary power law (e.g. Kara et al. 2013b; Cackett et al. 2014; Uttley et al. 2014; Chainakun \& Young 2015). In H 1743-322 the disk-to-power law flux ratio at 1 keV, i.e. at the energy where the soft lag starts to emerge in the lag-energy spectra, is $\sim 62\pm27$, $\sim 35\pm8$, $\sim 33\pm11$, $\sim 26\pm10$ percent, respectively for O1, O2, O3, and O4 (Table \ref{tab:fits} and Sect. \ref{sec:analysis}). We measured the disk fraction at this same energy also for GX 339-4 in the hard state, during the observation at $\rm{L_{3-10\ keV}/L_{Edd}\sim 0.004}$. To this aim we used the best fit model reported in DM15 and De Marco et al. (2015b). We estimated a disk-to-power law flux ratio of $\sim 30\pm10$ percent at 1 keV, thus consistent with the values derived for H 1743-322 within a factor $\sim$2. 
In the present case, dilution would reduce the intrinsic lag approximately by a factor $R/(1+R)$ (where $R$ here parametrizes the disk flux fraction relative to the power law, e.g Uttley et al. 2014). Therefore, a variation of a factor $\sim$ 2 of the disk fraction would produce a similar variation in the measured lag.
Thus, we conclude that dilution is a viable possibility to (at least in part) explain the observed offset of reverberation lag amplitude in H 1743-322 and GX 339-4 at the same luminosity in the hard state.\\
{\bf \emph{Inclination:}}
As reported in Sect. \ref{intro} several observational properties (e.g. X-ray dipping episodes, the shape of the HID, and the detection of accretion disk winds traced by high-ionization Fe K lines) of H 1743-322 point to the system being observed at high inclination ($i\sim 75^{\circ}$, Steiner et al. 2012). On the contrary, these same properties suggest GX 339-4 to be a low inclination system (DM15, see also Soria, Wu, \& Johnston 1999; Ludlam, Miller \& Cackett 2015).
Inclination has the effect of reducing the path length of photons reprocessed in the near side of the disk and increasing that of photons coming from the far side of the disk (e.g. Peterson 2001). However, the measured lag is the average of the lags associated with all the possible paths from the X-ray source to the disc and from the disc to the observer, weighted for the fraction of reprocessed flux (e.g. Wilkins \& Fabian 2013). Thus, quantifying the effects of inclination on the measured lag requires computation of the disk response functions from general relativistic ray-tracing simulations. Results of these simulations show that reverberation lags are slightly shorter in higher inclination systems (e.g. Emmanoulopoulos et al. 2014; Cackett et al. 2014). This is a consequence of the fact that the initial rise of the response function occurs at shorter time delays. Therefore, accounting for inclination effects would result in an increase of the observed offset between the lag measured in H 1743-322 and GX 339-4, though this increase is predicted to be small (of the order of 10 percent, e.g. Emmanoulopoulos et al. 2014; Cackett et al. 2014, thus smaller than the uncertainty on the lag measurement).\\
{\bf \emph{Intrinsic difference:}}
It is also plausible that the observed difference is intrinsic, thus indicating that the inner accretion flow geometry at a given luminosity in the hard state differs from source to source. Fits of the lag-energy spectra with self-consistent spectral-timing models are needed to confirm this hypothesis.

\section*{Acknowledgments}
The authors thank the anonymous referee for helpful comments. This work is based on observations obtained with XMM-{\it Newton}, an ESA science mission with instruments and contributions directly funded by ESA Member States and NASA. The authors thank F. Haberl, P. Predehl, C. Jin, and T. Dwelly for useful discussion and suggestions. G.P. acknowledges support via the Bundesministerium f\"ur Wirtschaft und Technologie/Deutsches Zentrum f\"ur Luft und Raumfahrt (BMWI/DLR, FKZ 50 OR 1408) and the Max Planck Society.



\end{document}